\documentclass[a4paper,12pt, epsfig]{article}  \pdfoutput=1
\usepackage{epsfig}
\usepackage{amssymb,latexsym}
\usepackage{amsfonts}
\usepackage{amsmath}

\newskip\humongous \humongous=0pt plus 1000pt minus 1000pt

\newif\ifdtup

\catcode`\@=11

\@addtoreset{equation}{section}
\def\theequation{\thesection.\arabic{equation}}

\def\@normalsize{\@setsize\normalsize{15pt}\xiipt\@xiipt
\abovedisplayskip 14pt plus3pt minus3pt%
\belowdisplayskip \abovedisplayskip
\abovedisplayshortskip \z@ plus3pt%
\belowdisplayshortskip 7pt plus3.5pt minus0pt}

\def\small{\@setsize\small{13.6pt}\xipt\@xipt
\abovedisplayskip 13pt plus3pt minus3pt%
\belowdisplayskip \abovedisplayskip
\abovedisplayshortskip \z@ plus3pt%
\belowdisplayshortskip 7pt plus3.5pt minus0pt
\def\@listi{\parsep 4.5pt plus 2pt minus 1pt
      \itemsep \parsep
      \topsep 9pt plus 3pt minus 3pt}}

\relax

\catcode`@=12

\catcode`\@=11

\def\section{\@startsection{section}{1}{\z@}{3.5ex plus 1ex minus
    .2ex}{2.3ex plus .2ex}{\large\bf}}

\def\thesection{\arabic{section}}
\def\thesubsection{\arabic{section}.\arabic{subsection}}

\def\appendix{\setcounter{section}{0}
  \def\thesection{Appendix \Alph{section}}
  \def\thesubsection{\Alph{section}.\arabic{subsection}}
  \def\theequation{\Alph{section}.\arabic{equation}}}

\def\SymBoxes#1#2#3#4{\newdimen\un@t \un@t#3%
\raisebox{#1}{\rule{#2\un@t}{#4}\hskip-#2\un@t
\@tempdimb\un@t \advance\@tempdimb by-#4\@tempcntb#2\relax%
\@whilenum{\@tempcntb>0}\do{
\rule{#4}{\un@t}\hskip\@tempdimb \advance\@tempcntb by\m@ne}%
\hskip-#2\un@t \rule[\un@t]{#2\un@t}{#4}%
\rule[\un@t]{#4}{#4}\hskip-#4
\rule{#4}{\un@t}}\hskip-#4}                

\begin{document}


\newcommand{\dd}{\textrm{d}}

\newcommand{\beq}{\begin{equation}}
\newcommand{\eeq}{\end{equation}}
\newcommand{\bea}{\begin{eqnarray}}
\newcommand{\eea}{\end{eqnarray}}
\newcommand{\beas}{\begin{eqnarray*}}
\newcommand{\eeas}{\end{eqnarray*}}
\newcommand{\defi}{\stackrel{\rm def}{=}}
\newcommand{\non}{\nonumber}
\newcommand{\bquo}{\begin{quote}}
\newcommand{\enqu}{\end{quote}}
\renewcommand{\(}{\begin{equation}}
\renewcommand{\)}{\end{equation}}
\def\de{\partial}
\def\Om{\ensuremath{\Omega}}
\def\Tr{ \hbox{\rm Tr}}
\def\rc{ \hbox{$r_{\rm c}$}}
\def\H{ \hbox{\rm H}}
\def\HE{ \hbox{$\rm H^{even}$}}
\def\HO{ \hbox{$\rm H^{odd}$}}
\def\HEO{ \hbox{$\rm H^{even/odd}$}}
\def\HOE{ \hbox{$\rm H^{odd/even}$}}
\def\HHO{ \hbox{$\rm H_H^{odd}$}}
\def\HHEO{ \hbox{$\rm H_H^{even/odd}$}}
\def\HHOE{ \hbox{$\rm H_H^{odd/even}$}}
\def\K{ \hbox{\rm K}}
\def\Im{ \hbox{\rm Im}}
\def\Ker{ \hbox{\rm Ker}}
\def\const{\hbox {\rm const.}}
\def\o{\over}
\def\im{\hbox{\rm Im}}
\def\re{\hbox{\rm Re}}
\def\bra{\langle}\def\ket{\rangle}
\def\Arg{\hbox {\rm Arg}}
\def\exo{\hbox {\rm exp}}
\def\diag{\hbox{\rm diag}}
\def\longvert{{\rule[-2mm]{0.1mm}{7mm}}\,}
\def\a{\alpha}
\def\b{\beta}
\def\e{\epsilon}
\def\l{\lambda}
\def\ol{{\overline{\lambda}}}
\def\ochi{{\overline{\chi}}}
\def\th{\theta}
\def\s{\sigma}
\def\oth{\overline{\theta}}
\def\ad{{\dot{\alpha}}}
\def\bd{{\dot{\beta}}}
\def\oD{\overline{D}}
\def\opsi{\overline{\psi}}
\def\dag{{}^{\dagger}}
\def\tq{{\widetilde q}}
\def\L{{\mathcal{L}}}
\def\p{{}^{\prime}}
\def\W{W}
\def\N{{\cal N}}
\def\hsp{,\hspace{.7cm}}
\def\bo{\ensuremath{\hat{b}_1}}
\def\bfo{\ensuremath{\hat{b}_4}}
\def\co{\ensuremath{\hat{c}_1}}
\def\cfo{\ensuremath{\hat{c}_4}}
\newcommand{\C}{\ensuremath{\mathbb C}}
\newcommand{\Z}{\ensuremath{\mathbb Z}}
\newcommand{\R}{\ensuremath{\mathbb R}}
\newcommand{\rp}{\ensuremath{\mathbb {RP}}}
\newcommand{\cp}{\ensuremath{\mathbb {CP}}}
\newcommand{\vac}{\ensuremath{|0\rangle}}
\newcommand{\vact}{\ensuremath{|00\rangle}                    }
\newcommand{\oc}{\ensuremath{\overline{c}}}

\newcommand{\Vol}{\textrm{Vol}}

\newcommand{\half}{\frac{1}{2}}

\def\changed#1{{\bf #1}}

\begin{titlepage}

\def\thefootnote{\fnsymbol{footnote}}

\begin{center}
{\large {\bf
Closed Timelike Curves in the Galileon Model
  } }

\bigskip

{\large
\noindent
Jarah Evslin${}^{1}$\footnote{\texttt{jarah@ihep.ac.cn}} and Taotao Qiu${}^{2}$\footnote{\texttt{xsjqiu@gmail.com}} }
\end{center}

\renewcommand{\thefootnote}{\arabic{footnote}}

\begin{center}
\vspace{0em}
{\em  { 
${}^{(1)}$TPCSF, IHEP, Chinese Acad. of Sciences, YuQuan Lu 19(B), Beijing 100039, China\\

\vspace{.2cm}

${}^{(2)}$Dept of Physics, Chung-Yuan Christian U, Chung-li 320, Taiwan, China\\

\vskip .4cm}}

\end{center}

\vspace{3.1cm}

\noindent
\begin{center} {\bf Abstract} \end{center}

\noindent
It has long been known that generic solutions to the nonlinear DGP and Galileon models admit superluminal propagation.  In this note we present a solution of these models which also admits closed timelike curves (CTCs).  We observe that CTCs only arise when, according to each observer, there exists some region in which the higher derivative terms are larger than the 2-derivative kinetic term.

\vfill

\begin{flushleft}
{\today}
\end{flushleft}
\end{titlepage}

\hfill{}


\setcounter{footnote}{0}

\section{Introduction}
Given a configuration on an initial spatial slice, a physical theory calculates the configuration at any moment in the future.  Not any initial condition is necessarily admissible, for example the requirement that a field be single-valued or that an observable be gauge-invariant may lead to constraints on the set of allowed initial conditions.  In popular theories of Nature these constraints may often be expressed locally, at least in Fourier space, and one may easily determine whether they are satisfied by a given set of initial conditions.  But what if it were only effectively possible to determine which initial conditions are consistent {\it{after}} solving the theory?  Such a theory would, at the very least, be very difficult if not impossible to quantize as no obvious candidate would present itself for a Hilbert space of states.

In this note we will be interested in field theories which develop closed timelike curves (CTCs).  We will demand that the fields be single-valued, which leads to a consistency constraint for each curve.  As each curve has a finite length, this consistency condition is necessary nonlocal.  In fact, as the length can be arbitrary in these theories, the condition is even nonlocal in Fourier space.  A sufficient condition for solving these constraints would be to restrict attention to configurations which never develop CTCs.  However we will argue that this condition is very nonlocally and nonlinearly encoded in the initial data, and so determining whether it is satisfied is no easier than solving for the entire future evolution of the system.

There is another means of escape from CTCs, the universe may conspire to eliminate them \cite{hawking}.  This will happen if the theory in question is only a low energy effective theory, and the full UV theory is chronologically protected.  A recent example of this phenomenon has been described in Ref.~\cite{Horava}.  In such cases, even at moments in which the energy scale is sufficiently low that one would ordinarily trust the low energy effective description, the UV theory may intervene and avoid the creation of CTCs, violating the low energy equations of motion.  Such a mechanism for protecting the chronology of k-essence theories has been evoked in Ref. \cite{Alexribut}.   Already Ref. \cite{Nima} contains examples of effective theories that develop CTCs within the domain where dimensional analysis suggests that the effective theory is to be trusted.  However, as will be the case below, different observers may disagree on when the effective theory is to be trusted, and it may not be obvious which observer is correct.  For example, in the case of gauge/gravity duality, in Ref. \cite{Benaabyss} Einstein gravity breaks down at distances which appear macroscopic from the viewpoint of a local observer, in that case it is instead the Planck length observed by a distant, asymptotic observer which determines the validity of the low energy effective theory.

In this short note we will consider the Galileon models \cite{Galileon}, which are higher-derivative theories of a single scalar field generalizing a conjectured short distance limit\footnote{It may be that the actual short distance limit contains additional couplings which modify an equation of motion and also add degress of freedom and another equation of motion \cite{nodecoup}.}  of the DGP model \cite{DGP}.  We will consider these models in flat Minkowski space, decoupled from gravity\footnote{A ghostfree coupling to gravity necessarily violates the Galilean symmetry \label{referee} and so one may expect the results below to be modified.  It would nonetheless be interesting, and essential for cosmological applications, to extend our results to this case.} , as many of the advantages and problems of these models already appear in this setting \cite{consistency}.    While the Lagrangian, equations of motion and background geometry are all Lorentz-invariant, general solutions of these models spontaneously break Lorentz symmetry.  Superluminal propagation and therefore CTCs only appear in these less symmetric backgrounds.  The fact that we will work in Minkowski space, decoupling gravity, has the great advantage that there are well-defined notions of time.  In fact, every vector which is timelike with respect to the Minkowski metric yields a reference frame and in each reference frame, up to an irrelevant shift, one may define an absolute notion of time.  In particular, a CTC must, during some open interval, travel backwards with respect to the time of any given inertial frame.  However, as we will see in the example below, different observers will in general not agree on just which interval this is.

\section{The Galileon Models}
In this section we will briefly review those aspects of the Galileon models \cite{Galileon} which will be relevant for our example.   The single scalar field $\phi$ is described by the Lorentz-invariant Lagrangian density
\beq
\L=c_1\L_1-\frac{1}{2}\partial_\mu\phi\partial^\mu\phi+c_3\L_3+c_4\L_4+c_5\L_5 \label{Lag}
\eeq
where $c_k$ are real parameters and the $\L_k$ are given in terms of $\Pi_{\mu\nu}=\partial_\mu\partial_\nu\phi$ by
\begin{eqnarray} 
{\cal L}_1 & = & \phi \label{lag1} . \label{nonlin}\\
{\cal L}_3 & = & - \frac{1}{2} \,  \Tr(\Pi) \, \partial \phi \cdot \partial \phi \nonumber\\
{\cal L}_4 & = & - \frac{1}{4} \big( \Tr(\Pi)^2)\,  \partial \phi \cdot \partial \phi - 2 \,  \Tr(\Pi) \, \partial \phi \cdot \Pi \cdot \partial \phi - \Tr(\Pi ^2) \, \partial \phi \cdot \partial \phi + 2 \, \partial \phi \cdot \Pi ^2 \cdot \partial\phi \big) 
\label{Galileo4} \nonumber\\
{\cal L}_5 & = & -\frac{1}{5} \big(
\Tr(\Pi)^3 \, \partial \phi \cdot \partial \phi- 3 \Tr(\Pi)^2 \,  \partial \phi \cdot \Pi \cdot \partial \phi
-3 \Tr(\Pi)\Tr( \Pi^2) \,  \partial \phi \cdot \partial \phi\nonumber\\&&+6 \Tr(\Pi) \, \partial \phi \cdot \Pi^2 \cdot \partial\phi
+2 \Tr(\Pi ^3) \, \partial \phi \cdot \partial \phi
+3 \Tr(\Pi ^2) \, \partial \phi \cdot \Pi \cdot \partial \phi 
- 6 \,  \partial \phi \cdot \Pi^3 \cdot \partial\phi \big) .\nonumber
\end{eqnarray} 
For simplicity in Section \ref{ctcsez} we will consider the special case $c_1=c_4=c_5=0$ which describes a limit of the 5-dimensional DGP model \cite{DGP}.   

Not only are the equations of motion second order, but the variation of each term individually can be written as a sum of products of tensors built from the derivatives of $\phi$, and each summand contains at least one trace of a positive power of the matrix $\partial_\mu\partial_\nu\phi$.  Therefore if all powers of this matrix are traceless for a particular configuration of $\phi$, for example if it is nilpotent, then the equations of motion will be satisfied for any value of the constants $c_k$.

In Ref. \cite{consistency} a solution of the equations of motion was found for a configuration coupled to an external source, treated as a stationary delta function of the trace of the stress tensor.  Small perturbations about this solution propagate superluminally \cite{Nima}.  In fact, superluminal propagation is generic in these solutions, as was shown in Ref. \cite{positivity}.   In that reference the authors suggested removing the superluminal propagation by reiterating that the Galileon theory itself is not a consistent theory, but rather may only be interpreted as a low energy effective theory.  Furthermore, they found that the high energy theory becomes relevant whenever the nonlinearities of the Galileon theory are relevant.  In other words, the only terms that one may trust in the Galileon theory are, in this interpretation, the linear terms.  These terms do not respect a Galilean symmetry and do not share the features of the Galileon theory which make it attractive for model building.  For example, the stable, null-energy condition violating solutions are excluded.  In effect, this prescription, removing the nonlinearities of the Galileon theory, destroys all of the theories attractive features.  In Ref.~\cite{genesis} the authors go further and claim that, "If we decide to ban superluminality from our effective theory, we also lose predictivity for cosmological observables."  In Ref.~\cite{bigal,bigal2} the authors propose a new theory which, at least for some solutions, avoids superluminality.  However the genericity of this feature remains to be seen.

We would like to claim that the aforementioned draconian modification of the Galileon theory is premature.  As has been stressed in Ref. \cite{Alexribut} and in this context in Ref.~\cite{genesis}, superluminal propagation itself is not a problem.  A problem only occurs when it can be used to create CTCs, which according to the general logic in Ref. \cite{Nima} requires two natural, very different, local reference frames.  The creation of such frames without, for example, introducing tachyons, is in general nontrivial.  In fact, even if the theory admits solutions with CTCs, it is still not automatically unhealthy.  Even general relativity on a timelike circle contains CTCs.  The question is whether an easily implemented prescription for choosing initial conditions and solving the Cauchy problem exists such that these CTCs do not lead to an inconsistency such as a multiply defined field.  On the other hand, if internal creatures can make time machines from the kinds of configurations that cannot be excluded with local conditions on the initial conditions, then there are nonlocal constraints and the theory is not likely to by quantizable.  The theory may nonetheless still be valid as an effective theory, so long as each term in the Lagrangian is applicable at least in some configurations.  After all, some authors have even claimed that Einstein gravity itself when coupled to idealized nonrelativistic matter allows the creation of CTCs \cite{timemachine}.

We will show in the next section that the Galileon theory indeed does admit configurations with CTCs, and claim that the predecessors of these configurations have no obvious pathologies.  However, in all of these configurations, from the point of view of any observer, there exists a region where the higher derivative terms dominate the 2-derivative terms in the Lagrangian and so, if the Galileon theory is viewed as an effective theory, then these configurations will be beyond the regime in which dimensional analysis suggests that it may be trusted.  On the other hand our example demonstrates that if one considers the Galileon theory alone as a UV complete theory, then one is confronted with nonlocal constraints.  As one of the main advantages of the Galileon theory is that Lorentz-invariance is only spontaneously broken and so is restored in the UV, which allows it to be coupled to gravity, the UV incompleteness of the Galileon theory limits its utility.

The perhaps more surprising fact is that different observers disagree on just where the solution is beyond this regime of validity, and a discrete symmetry of the configuration relates these different candidate locations.  Therefore the effective theory must, as in Refs. \cite{Benaabyss,Nima} break down even when dimensional analysis leads one to believe that it is reliable.  Of course, it would be of great interest to know whether Einstein gravity shares the same behavior, for example rendering it invalid as an effective theory in very redshifted locations such as near the horizon of a black hole or in the distant past.

\section{A Galileon Configuration with a CTC} \label{ctcsez}

Finally we will describe our example of a configuration of the Galileon theory with no obvious pathologies that nonetheless evolves to a configuration with CTCs.  As the equations of motion are of second order, given a value of $\phi$ and its first time derivative on a Cauchy surface, one may evolve the configuration forwards or backwards using the equations of motion, at least up to constraints caused by CTCs.  Therefore it will be sufficient to find a solution during a brief period of time which manifests CTCs, and then say in words that while CTCs prevent the propagation of small fluctuations, the solution itself may be uniquely evolved backwards in time by the equations of motion.  The form of the solution and the equations of motion is such that the features apparently become more dilute as the solution is propagated backwards; the higher derivative terms become subdominant.  Therefore the initial configuration, a configuration arising a bit earlier than the formation of the CTCs, appears difficult to exclude using any local selection criteria.  

Following the basic strategy proposed in Ref.~\cite{strat} and in this context in Ref.~\cite{Nima}, we will construct a configuration with closed timelike curves from multiple regions with superluminal propagation traveling with different velocities.  At the moment in which the closed timelike curve exists, each region will be a cylindrical bar.  We will begin by describing a single such bar.

\subsection{A single bar}

We will consider a bar immersed in a background with a constant value of $\phi$.  For simplicity, we will set $\phi=0$ outside of the bar.  At time $t=0$,  the bar is extended in the $x$ direction from $x=-A$ on the left to $x=A$ on the right, while it is a disc in the other directions $y^2+z^2\leq B^2$.  We will be interested in a signal propagating along the axis $y=z=0$ of the bar, and so the dependence upon $r=\sqrt{y^2+z^2}$ will simply be an interpolation between the value of $\phi$ at $r=0$ and $\phi=0$ at $r=B$ such that the second $r$ derivative is continuous and in particular vanishes at $r=0$ and $r=B$.  As there is a single equation of motion to satisfy, the interpolating function can essentially be chosen at will, different choices simply lead to different profiles of the second time derivative of $\phi$ and so affect the configuration in the past and future.  One choice at the moment $t=0$, which satisfies all of these conditions, is
\beq
\phi(x,y,z,t)|_{t=0}=\left(1-\frac{10r^3}{B^3}+\frac{15r^4}{B^4}-\frac{6r^5}{B^5}\right)\phi(x,t=0). \label{buono}
\eeq
Insofar as this configuration can result from seemingly benign initial conditions, this time-dependence is irrelevant as our CTC will be localized at $t\sim 0$. The vanishing second derivatives in the $y$ and $z$ directions at $r=0$ and $r=B$ guarantees that continuity of the solution for small $t$, and also that the $r$-dependence will not affect the superluminal propagation of the small fluctuations at $r=0$ and $t\sim 0$ which will be discussed below.





The causal properties of our curve $y=z=0$ then depend entirely on the $x$ and $t$ dependence of $\phi$ at $r=0$.   We will restrict our attention to the propagation of small perturbations of the Galileon field
\beq
\phi=\phi_0+\delta\phi. \label{pdicomp}
\eeq
The decomposition (\ref{pdicomp}) leads to a decomposition of the Lagrangian  (\ref{Lag}). 
The terms which are independent of $\delta\phi$ are not relevant for the propagation of these perturbations.  The terms which are linear automatically vanish due to the equations of motion.  The terms which are cubic or higher are negligible as we are considering small fluctuations.

Therefore, mirroring the analysis of Ref. \cite{positivity} in the conformal Galileon case, we are left with the quadratic terms  
\beq
\delta\L=-\frac{1}{2}(\partial_\mu\phi)^2(1+2c_3\Box\phi_0)+c_3(\partial^\mu\partial^\nu\phi_0)\partial_\mu\delta\phi\partial_\nu\delta\phi
=-\frac{1}{2}G^{\mu\nu}\partial_\mu\delta\phi\partial_\nu\delta\phi
\eeq
where $G$ is the inverse metric which describes the causal propagation of small fluctuations.  The constant $c_3$ may be absorbed by an overall rescaling of $\phi$.   Below we will be interested in solutions in which $\phi_0$ is harmonic, and so
\beq
G^{\mu\nu}=\eta^{\mu\nu}-\partial^\mu\partial^\nu\phi_0
\eeq
where $\eta^{\mu\nu}$  is the inverse Minkowski metric.

As a first attempt, let us consider a field configuration which moves left at the speed of light
\beq
\phi_0|_{r=0}=f(x+t)
\eeq
where $f$ is an arbitrary function.   This function is harmonic, and also it satisfies the equations of motion (for any values of the constants $c_k$) since the matrix $\partial_\mu\partial_\nu\phi_0$ is traceless (using the inverse Minkowski metric $\eta^{\mu\nu}$) and nilpotent of order two.  Notice that, unlike the superluminal solutions in Refs.~{\cite{superlumdgp,positivity}}, our solution requires no external stress tensor and is therefore a generic feature of all Galileon models.  Defining $h$ to be minus the second derivative of the function $f$ with respect to its argument
\beq
h(w)=-f^{\prime\prime}(w)
\eeq
the inverse metric for perturbations is simply
\beq
G^{\mu\nu}= \left(
\begin{array}{cc}
  -1+h & h \\
 h & 1+h
\end{array}
\right)
\eeq
which can be inverted to yield the metric
\beq
G_{\mu\nu}= \left(
\begin{array}{cc}
  -1-h & h \\
 h & 1-h
\end{array}
\right) .
\eeq

This metric describes the propagation of small oscillations of the $\phi$ field.  On the $t-x$ plane there are, up to a constant factor, two null vectors with respect to this metric
\beq
v_1= \left(
\begin{array}{cc}
  1 \\
 1 
\end{array}
\right)
\hsp
v_2= \left(
\begin{array}{cc}
  1-h \\
 -h-1
\end{array}
\right).
\eeq
These vectors lie on the null light cone, bounding the future light cone.  The first vector, $v_1$, lies along the usual lightcone, followed by particles traveling at the speed of light.  This means that oscillations moving to the right, in the opposite direction to the bar, travel at precisely the speed of light.    The fact that the speed of these signals is independent of the value of $h$ in the rod will be exploited momentarily.

The second vector,  $v_2$, is more interesting.  It is superluminal precisely when $h>0$.  As we have described, a CTC requires that, in any given frame, part of the curve must move backwards in time.  In the frame given by the coordinates that we have chosen, this implies that $h>1$, which is just the condition that the higher derivative $c_3$ terms of the Lagrangian (\ref{nonlin}) be greater than the two derivative term, in which case the Galileon theory should not be trusted if it is an effective theory, but should be if it is considered as the full UV theory.  In this case fluctuations travel backwards in time in the rest frame that we are describing, but only left-moving particles.  Thus if one could build a left-moving rod with $h>1$ everywhere, then a left-moving signal would be able to traverse the rod instantaneously, even arriving before it left, and such rods could then be pieced together to form a CTC.


Needless to say, $h(x+t)$ is not an arbitrary function as outside of the rod the Galileon field vanishes
\beq
\phi(\pm A)=\partial_x \phi(\pm A)=0. \label{bordo}
\eeq
This apparently yields a serious problem, as if one demands that $f=f\p=0$ at the boundaries then the integral of $h$ must vanish.  This would imply a large region of negative $h$, in which the propagation is subluminal and therefore a left-moving trajectory in fact moves appreciably forward in time.  Thus a CTC cannot be constructed entirely of such components.

The solution to this problem is to recall that the speed of right-moving signals are unaffected by left-moving rods.  A spatial reflection of the above calculation also indicates that our left-moving signal would be unaffected by a right-moving rod.  Therefore one may introduce a right-moving rod which enforces the boundary condition (\ref{bordo}) and which our left-moving signal may traverse not instantaneously, but nonetheless at the speed of light.  The speed of the left-moving signal in the right-moving rod is always the speed of light, independently of the value of $h$ in the right-moving rod.   Therefore $h$ in the right-moving rod may be taken to be as negative as one likes, so that the rod may be as short as one likes and the time lost traversing it will therefore be arbitrarily short.  

Thus such a right-moving rod may be added as desired to the configuration to enforce the boundary condition.  For example, a right-moving rod extending from $x=-\epsilon$ to $x=\epsilon$ with $h$ equal to $-A/\epsilon$ times its' value in the left-moving rod.  At $t=0$, the $\phi$ field is continuous at the interface between the two rods, only its third derivative diverges which does not appear in the equations of motion.   Therefore no new pathology is introduced around $t=0$.  However, using the equations of motion to evolve such a configuration back to earlier times is likely to result in rather large derivatives as the rods collide.  This reinforces the observation that, at least for this class of configurations, solutions with CTCs are beyond the validity of the derivative expansion.

\subsection{Robustness of the cylinder solution}

While we have seen that our configurations are beyond the validity of the derivative expansion, this does not necessarily imply a pathology.  In this section we will discuss three potential problems that may worry the reader.  We will in turn discuss the boundary conditions of the cylindrical solution, we will discuss its local stability and finally we will apply the analysis of Ref.~\cite{consistency} to examine the consistency of the quantum perturbative expansion.

\noindent
{\bf{Boundary conditions}}

The solid cylinder has three boundaries.  One is a hollow cylinder at $r=B$ which extends from $x=-A$ to $x=A$.  The other two are discs $r\leq B$ at $x=\pm A$.  We have already described, in words, the boundary conditions on the cylinder.  For concreteness we will consider a subclass of the possible configurations, which however includes (\ref{buono}) as an example
\beq
\phi(x,y,z,t=0)=g(r)f(x).  \label{fattore}
\eeq
Our $r=B$ boundary condition can then be written
\beq
g^{\prime\prime}(r)|_{r=B}=g\p(r)|_{r=B}=g(B)=0 \label{gcond}
\eeq
which as desired is satisfied by (\ref{buono}).  

Galilean theories have second order equations of motion, and since $c_1=0$ the {\it{source free}} equations of motion can be written in the form
\beq
(\Box\phi)(1+c_3\Box\phi)-c_3 (\partial_{\mu}\partial_{\nu}\phi)(\partial^{\mu}\partial^{\nu}\phi)=0.
\eeq
This implies that the equations of motion are satisfied whenever $\Box\phi=(\partial_{\mu}\partial_{\nu}\phi)(\partial^{\mu}\partial^{\nu}\phi)=0$.  The boundary conditions (\ref{gcond}) imply that all of the second derivatives of $\phi$ vanish separately.  Therefore  the equations of motion are satisfied at the boundary with no source.  This is in contrast to the stationary superluminal solutions considered in Ref.~\cite{positivity}, which are the result of the backreaction on the Galileon of a source with a nontrivial external stress tensor.  Here the only contribution to the stress tensor arises from the Galileon itself, which is the only field in this model, and even this contribution to the stress tensor vanishes at the cylindrical boundary as the second derivatives of the field vanish.

Thus no stress tensor source is required at the cylindrical boundary $r=B$.  What about the discs at $x=\pm A$?  One could similarly force the second derivatives to vanish independently by imposing that the second derivative of $f$ vanishes at the boundaries.  However this is not necessary, the boundary conditions mentioned above
\beq
f\p(x)|_{x=\pm A}=f(\pm A)=0 \label{disc}
\eeq
are sufficient to impose $\Box\phi=(\partial_{\mu}\partial_{\nu}\phi)(\partial^{\mu}\partial^{\nu}\phi)=0$ and so to satisfy the vacuum equations of motion.  To see this, note first that all second derivatives of $\phi$ in which at least one derivative is in the $y$ or $z$ direction will either contain zero or one $x$ derivatives, in which case the second or first condition in (\ref{disc}) will impose that the corresponding derivative vanishes.  The second $x$ and $t$ derivatives do not vanish, however, as $f$ only depends upon them in the combination $x+t$ and our metric is Minkowski, they cancel.  Again no external source is necessary.  

\noindent
{\bf{Instabilities}}

While the stress tensor vanishes at the boundaries of the cylinder, one may nonetheless be concerned that the null energy condition may be violated inside of the cylinder, which may lead to local instabilities.  Of course, one of the strengths of the Galileon theories is that they possess configurations which stably violate the null energy condition \cite{Galileon}, but there are also configurations which violate the null energy condition and are unstable.   We will consider two of the most serious instabilities, ghost instabilities and exponential instabilities characterized by a negative speed of sound squared.  These instabilities are serious problems because the highest frequency modes are the most unstable, and so the decay occurs instantaneously.  If our configuration develops such instabilities before developing CTCs, then one may question the relevance of these solutions. 

These are both instabilities with respect to small fluctuations in the Galileon field.  Therefore they are captured by the small fluctuation expansion, and in particular by the metric for perturbations.  Ghost instabilities occur if both of the metric's eigenvalues change sign, whereas the exponential instability occurs if only one changes sign.  The determinant of the metric at $r=0$ is easily seen to be equal to $-1$.  This is the product of the eigenvalues, therefore an eigenavalue may change sign only if it passes zero, which implies that the other is divergent.  However the sum of the eigenvalues, given by the trace of the the metric, is equal to $-2h$ which is finite for our solutions.  Therefore the eigenvalues do not change sign and no such local instability is expected at $r=0$.  Of course, once the closed timelike curves have been created, these may lead to global instabilities, but such problems lie beyond the scope of this note.

\noindent
{\bf{Time dependence}}

We have shown that an infinite planewave is an exact solution of the Galilean equations of motion.  However an infinite planewave would not have closed timelike curves.  To obtain closed timelike curves we have introduced an $r$-dependence, which at $t=0$ is described by the simple Ansatz (\ref{fattore}).  The price for this $r$-dependence is that the first and second derivatives of $\phi$ in the $r$-direction are nonzero.  These derivatives affect the equations of motion, and as we have described above, we choose the second time derivative of $\phi$ to solve these corrected equations of motion.  

This means that $\phi$ is no longer a planewave, but deforms in time.  The $(x-t)$-dependence is only a leading order behavior, providing also an initial condition for the first time derivative of $\phi$ which is used for integrating the equations of motion.  The resulting correction to the second time derivatives means that, upon integrating the equations of motion, there will be terms proportional to $g\p(r)t^2$ and also to $g^{\prime\prime}(r)t^2$.  These terms of course affect the effective metric, and so do our CTC's exist after all?

As will be described in Subsec.~\ref{tanti}, we will construct a closed timelike curve which is contained entirely in an interval of time $t\in [-\epsilon,\epsilon]$ where $\epsilon$ can be made arbitrarily small by tuning the number of cylinders and their lengths.   Therefore it is sufficient that the solution at $r=0$, along with its first two derivatives, be unaffected during this short time interval.  

This can be arranged.  The key observation is that propagation in the $r$ direction is not superluminal in the planewave solution, it is in fact highly subluminal with corrections proportional to $g\p(r)$ and $g^{\prime\prime}(r)$.  Therefore if one simply declares that
\beq
g\p(r)=0\rm{\ \ \ for\ \ \ }r<\epsilon
\eeq
then the solution will, at $r=0$, be unaltered during the time interval $t\in [-\epsilon,\epsilon]$.  Therefore, although in the past and the future the rods will deform beyond recognition, this deformation does not occur during the time relevant for the CTC.  Of course, this deformation makes life complicated for whoever wishes to assemble the cylinders, he must solve the equations of motion backwards to find the appropriate initial conditions, but the solution to this difficult computation is beyond the scope of our note.

\noindent
{\bf{Quantum corrections}}

While we are in part motivated by a desire to understand whether the Galileon theory admits a quantum completion, our analysis itself is entirely classical.  Our result is a purely classical statement, that the classical solutions of a classical field theory develop closed timelike curves even when the initial conditions are apparently benign.  In particular our result is independent of whether the Galileon field is an elementary field that may be quantized, or a collective excitation such as a phonon that need not be quantized directly.  In Ref.~\cite{consistency} the authors consider a direct quantization of the Galileon field, or more precisely the DGP truncation which is equivalent to the theory discussed in this note.  

In our case the analysis is one step shorter, as our Galileon $\phi$ is already canonically normalized.  It has dimensions of inverse length.  Therefore the additional length scale $L_Q$ which, according to Ref.~\cite{consistency}, enters in the quantum theory is simply our coupling constant
\beq
L_Q^3=c_3.
\eeq
The authors claim that, as a consequence of Lorentz-invariance, this scale only enters into physical quantities in the combination
\beq
\alpha_q=L_Q^2\partial^2=c_3^{2/3}\partial^2
\eeq
where $\partial$ is the derivative scale $\sim\sqrt{(\partial^2\phi)/\phi}$.  In our case there are two such scales, in the two directions, $1/A$ and $1/B$.  Therefore we find two expansion parameters
\beq
\alpha_q^{(1)}=\frac{c_3^{2/3}}{A^2}\hsp
\alpha_q^{(2)}=\frac{c_3^{2/3}}{B^2}.
\eeq
The necessary condition for convergence in the sense of an asymptotic series is then simply that
\beq
c_3<<A^3,\ B^3.
\eeq
Thus for any value of $c_3$ the quantum perturbation theory is not sensible if either $A$ or $B$ is too small.  In the case of the free theory, $c_3=0$, the theory is always sensible but there are no CTC's.

\subsection{Assembling rods to make a configuration with CTCs} \label{tanti}

We have now constructed the basic ingredient in our configuration.  At a fixed moment in time, $t=0$, we have found a solution to the equations of motion which is a cylindrical rod.  The rod moves to the left at the speed of light, and a small perturbation in the $\phi$-field may traverse the rod in the left-moving direction instantaneously, if desired even arriving before it leaves.  The solution is beyond the validity of the derivative expansion, as the $3$-derivative terms in the action are as large as the $2$-derivative terms precisely when $h\geq 1$ and this is the condition under which the oscillation arrives instantaneously.  

As described in Ref.~\cite{Alexribut}, a complication in any such construction is that the equations of motion demand that the rod itself will deform and grow in time.  Indeed, using the equations of motion to run the solution either forwards or backwards in time, it seems likely that the rod will expand and smear out.  This needs to be checked, and we hope to return to this point in the future.  However, if indeed it is the case, then the initial conditions which lead to such rods will, while very fine-tuned, have small derivatives and so be difficult to exclude based on any local principle.  Thus reasonable seeming initial-conditions, which solve the constraints, are likely to lead to configurations with rods which we will soon argue have CTCs.  Our rods have an additional complication as, embedded in their cores, is a rod moving in the opposite direction.  Therefore these solutions are utterly deformed shortly before and after the CTC, however the CTC itself does not involve these temporal regions and so they are only relevant to the secondary question of understanding which initial conditions lead to CTCs.

\begin{figure}
\begin{center}
\includegraphics[scale=.50]{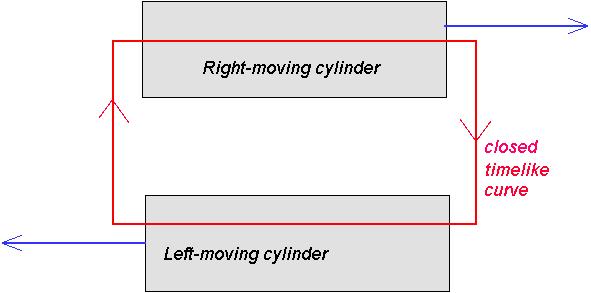}
\caption{Two rods pass each other with a finite impact parameter.  A potential CTC threads the two rods along the direction of their motion, traveling backwards in time as it travels through each.  However during the time required to travel between the rods, the inter-rod interactions are likely to destroy the configuration.}
\label{rc}
\end{center}
\end{figure}

Finally we are ready to assemble these rods to construct a configuration with CTCs.  The simplest possibility, mirroring the proposal in Ref.~\cite{Nima}, which proposed the extension of their construction to this model as an open problem\footnote{The existence of CTCs in the Galileon model was very recently posed as an open problem in Ref.~\cite{genesis}.}, would be to consider two rods traveling in opposite directions with an impact parameter $2B$ as seen in Fig.~\ref{rc}.

In this configuration, a small perturbation must twice travel between the cores, a distance of $2B$, at the speed of light.   This trip takes time $2B$.  In that time, given that the equations of motion are relativistic, on dimensional grounds one expects the rods themselves to deform with a characteristic distance scale of order $B$.  However this scale is the same scale as the impact parameter itself, therefore there is no justification for treating the two rod solutions independently, nonlinear couplings will generically alter them beyond recognition during this travel time, and so the single rod analysis will be invalid.  This is essentially a manifestation of the bubble expansion argument presented in Ref.~\cite{Alexribut}.

\begin{figure}
\begin{center}
\includegraphics[scale=.50]{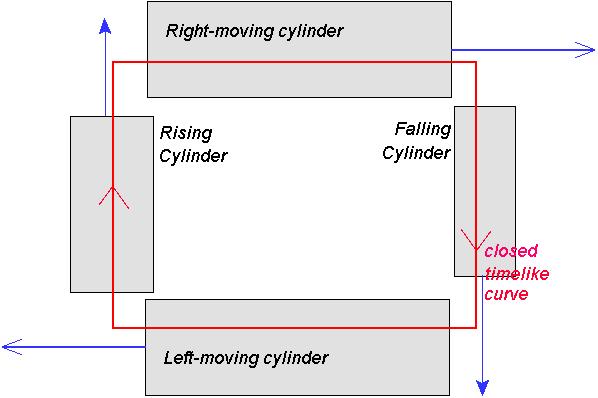}
\caption{A configuration is considered in which $n$ rods form a clockwise-moving $n$-gon.  Now, for a fixed rod radius, the time required by the perturbation to travel from the core of one rod to the next may, for some value of $n$ be arbitrarily small.  Therefore the interactions between the rods during this time interval may be neglected.}
\label{rc2}
\end{center}
\end{figure}

The origin of the problem is that, since outside of the rods the perturbation travels at the speed of light, the more one attempts to isolate the rods the more time one needs to wait for the signal and so the more time the rods have to distort each other.  We propose the following solution to this problem.  Instead of two rods moving in opposite directions, we will consider a polygon of rods all moving clockwise.  The case of four rods is depicted in Fig.~\ref{rc2}.


The advantage of such a configuration is that for any fixed rod radius $B$ the exit of the core of one rod may be arbitrarily close to the entrance of the next.  Therefore the amount of time in which the signal moves at the speed of light may be arbitrarily small and so the nonlinear interactions between the rods may be as suppressed as desired.  One may then choose $h$ to be large enough that the time gained traversing each rod is greater than that lost traveling between the rods, and a clockwise loop threading through all of the rods will describe a closed timelike curve.

An observer who is very boosted in the left-moving direction, which is perhaps a more natural local observer given that the bar itself is moving left at the speed of light, would claim that inside of his bar these oscillations move superluminally but not backwards in time.  He would also claim that it is the two-derivative term which dominates the Lagrangian, and so as an effective theory the Galileon theory should be trusted.  However he would claim that the other 3 images of his location, under the 90 degree rotation symmetry, indeed do exhibit time-inverted propagation and that by dimensional analysis the effective theory should not be trusted there.  Needless to say, this observer and his 4 images under the rotation symmetry do not agree on where the effective theory breaks down and so by symmetry they must all be wrong.  Despite the fact that in their reference frame a naive dimensional analysis suggests that the effective theory is reliable, it is not.  If the UV theory is the Galileon theory then there are CTCs, if it is a chronology protected theory, then the effective theory breaks down prematurely.


Evolving the configuration backwards in time, after an initial collision, the rods become separated by a larger distance and so we expect that they become more defuse, with a smaller second derivative.  Therefore an initial, earlier configuration, does not contain CTCs and likely lies further within the regime of validity of the effective theory.  Thus reasonable initial conditions lead to CTCs.  One may wonder whether this large $n$-gon may simply be replaced by a cleaner solution of a rotating ring, in which case no longer needs to add rods moving in the opposite direction.

\section* {Acknowledgement}

\noindent
JE is supported by the Chinese Academy of Sciences Fellowship for Young International Scientists grant number 2010Y2JA01. TQ is funded in parts by the National Science Council of R.O.C. under Grant No. NSC99-2112-M-033-005-MY3 and No. NSC99-2811-M-033-008 and by the National Center for Theoretical Sciences.   We are pleased to than D. Anselmi, Y. Cai, M. Li and X. Zhang for invaluable discussions.


\end{document}